\documentclass[conference]{IEEEtran}

\ifCLASSINFOpdf

\else

\fi

\usepackage{amsfonts}   
\usepackage{amsbsy} 
\usepackage{amsmath}
\usepackage{amssymb}
\usepackage{amsthm}

\usepackage{graphicx}
\usepackage{float}
\usepackage{subfloat}
\usepackage{subfig}
\usepackage[dvipsnames]{xcolor}

\usepackage{algorithm}
\usepackage{algorithmic}

\usepackage{tabularx}

\newtheorem{lemma}{Lemma}

\usepackage{amssymb}
\IEEEoverridecommandlockouts
\begin{document}

\title{Robust and Constrained Estimation of State-Space Models: A Majorization-Minimization Approach}

\author{\IEEEauthorblockN{Yifan Yu,
Shengjie Xiu, and Daniel P. Palomar \thanks{This work was supported by the Hong Kong GRF 16206123 research grant. The authors are with the Hong Kong University of Science and Technology (HKUST), Clear Water Bay, Kowloon, Hong Kong (e-mail: yyuco@connect.ust.hk; sxiu@connect.ust.hk; palomar@ust.hk)}\thanks{This work has been accepted by and presented at The Asilomar Conference on Signals, Systems, and Computers, Oct. 2024.}} 
\IEEEauthorblockA{Department of Electronic and Computer Engineering\\
The Hong Kong University of Science and Technology, Hong Kong}}

\maketitle

\begin{abstract}
In this paper, we present a novel optimization algorithm designed specifically for estimating state-space models to deal with heavy-tailed measurement noise and constraints. Our algorithm addresses two significant limitations found in existing approaches: susceptibility to measurement noise outliers and difficulties in incorporating constraints into state estimation. By formulating constrained state estimation as an optimization problem and employing the Majorization-Minimization (MM) approach, our framework provides a unified solution that enhances the robustness of the Kalman filter. Experimental results demonstrate high accuracy and computational efficiency achieved by our proposed approach, establishing it as a promising solution for robust and constrained state estimation in real-world applications.
\end{abstract}

\IEEEpeerreviewmaketitle

\section{Introduction}
State-space models capture system dynamics and establish connections between measurements and hidden states. They have found widespread applications in robotics, navigation, and economics. The Kalman filter, a well-known solution for state estimation in linear Gaussian systems, has demonstrated efficiency in estimating hidden state values. However, the Kalman filter faces limitations when applied to real-world scenarios. One limitation is its susceptibility to outliers, such as signals from unreliable sensors or financial market anomalies, which are often modeled using the Student-t distribution \cite{franses2008simple, zhu2013variational}. These outliers can significantly degrade the performance of the Kalman filter by violating its Gaussian noise assumptions. Additionally, incorporating additional constraints on state variables presents another challenge. In environmental monitoring, for example, ensuring adherence to physical laws or environmental constraints is crucial, but the Kalman filter lacks direct support for such constraints \cite{teixeira2009state,kyriakides2005multiple}.

Consequently, the development of robust estimation techniques for state-space models with Student-t distributed measurement noise and practical constraints becomes imperative in various practical applications. We encounter two limitations in existing work.

\emph{Limitation 1: Existing algorithms struggle to achieve both high accuracy and efficiency.} The particle filter is one commonly used method for robust filtering, approximating the posterior distribution through sequential sampling of particles. It can handle a wide range of heavy-tailed measurement noise distributions \cite{arulampalam2002tutorial,li2006t,xu2013robust}. However, the particle filter's computational complexity grows exponentially with the number of particles, leading to significant computational burden. Alternatively, Variational Bayes (VB) and Expectation Propagation (EP) methods provide computationally efficient alternatives by approximating the posterior distribution with certain assumptions. However, their approximation inaccuracies can limit performance \cite{bishop2006pattern}. Another approach involves optimization-based methods. For instance, the state estimation of Laplace filter is formulated an optimization problem \cite{wang2017laplace}. However, it is worth noting that this method has not yet been applied to the case of Student-t distributed measurement noise with constraints.

\emph{Limitation 2: Existing algorithms lack uniformity in effectively handling general constraints on the hidden state.}  Estimating states that must satisfy diverse constraints with varying structures and complexity poses challenges, leading to the development of separate models to address different constraint types. For linear equality constraints, reduction methods and perfect measurement techniques have been proposed \cite{wen1992model,porrill1988optimal}. Both of them incorporate the constraints into the model formulation. In the case of inequality constraints, the state estimate projection technique has been employed \cite{simon2002kalman}. This technique projects the unconstrained state estimation values onto constrained sets. In the context of particle filters, projection/rejection methods are applied to particles \cite{amor2018constrained}. However, these approaches encounter challenges related to convergence and accuracy. When dealing with more complex constraints, such as nonconvex constraints \cite{straka2012truncation,benidis2018optimization}, the existing algorithms lack a unified framework to effectively handle them. This poses further difficulties in achieving consistent and accurate state estimation.

In this paper, we introduce a novel optimization approach designed to efficiently estimate state-space models with Student-t noise and constraints. Our proposed approach addresses the challenge of dealing with the nonconvexity in Student-t formulation and various types of constraints by employing an MM method. By reformulating the constrained state estimation as an optimization problem, our approach offers a unified solution that encompasses and extends existing works for more general purposes. Through rigorous experimentation, we provide evidence of the high computational efficiency achieved by our proposed approach.

\section{Problem Formulation}

In this section, we present the formulation of the robust state-space model with the Student-t noise distribution. We then formally define the target constrained state estimation problem, while providing examples of the practical constraints.

\subsection{Robust State-space model with Student-t noise}
Firstly, we introduce the notation and related background. We consider a discrete-time linear state-space model with hidden state vector $\mathbf{x}_{k} \in \mathcal{R}^{n_x}$ and measurement vector  $\mathbf{y}_{k}  \in \mathcal{R}^{n_y}$ , where the time step is denoted as $k = 1,\dots, T$. The state-space model is defined as follows:
\begin{equation}
\begin{aligned}\mathbf{x}_{k+1}=\mathbf{A}\mathbf{x}_{k}+\mathbf{w}_{k}, &  & \quad & \mathbf{w}_{k}\sim\mathcal{N}\left(0,\mathbf{Q}\right);\\
\mathbf{y}_{k}=\mathbf{C}\mathbf{x}_{k}+\mathbf{v}_{k}, &  & \quad & \mathbf{v}_{k}\sim\prod_{i=1}^{n_{y}}\mathcal{T}\left(0,\sigma_{i},\nu_{i}\right),
\end{aligned}
\label{prob}
\end{equation}
where $\mathbf{w}_{k}\in \mathcal{R}^{n_x}$ represents the zero-mean Gaussian process noise with covariance matrix $\mathbf{Q}\in \mathcal{R}^{n_x \times n_x}$. $\mathbf{A} \in \mathcal{R}^{n_x \times n_x}$ and $\mathbf{C} \in \mathcal{R}^{n_x \times n_y}$ are the measurement and state transition matrix. The each element of measurement noise $\mathbf{v}_{k}\in \mathcal{R}^{n_y}$ is modeled as independent univariate Student-t noise, denoted by $\mathcal{T}\left(0,\sigma_i, \nu_i\right)$ with the following probability density function:
\begin{equation}
    p(v_{k,i}) = \frac{\Gamma\left(\frac{\nu_i+1}{2}\right)}{\sqrt{\pi\nu_i}\Gamma\left(\frac{\nu_i}{2}\right)}\left(1+\frac{v_{k,i}^2}{\nu_i\sigma_i^2}\right)^{-(\nu_i+1)/2},
\end{equation}
where $\sigma_i$ controls the scale, and 
$\nu_i$ is the degree-of-freedom. And we assume the initial state $\mathbf{x}_1$ follows Gaussian distribution $\mathcal{N}(\hat{x}_0, \mathbf{P}_0)$ .In the following sections, we denote all system parameters as $\boldsymbol{\theta} = \left\{\mathbf{A}, \mathbf{C},\mathbf{Q},\sigma_{1:n_y}, \nu_{1:n_y},\hat{x}_0,\mathbf{P}_0\right\}$.

\subsection{Constrained estimation problem\label{sec:II.B}}

In the context of constrained estimation, we aim to estimate the marginal distributions of hidden states $\mathbf{x}_k$ based on the available measurements $\mathbf{y}_k$ up to time $k$, while incorporating additional constraints on the estimated results.

To formulate the problem as an optimization problem, we consider the constrained Maximum a Posteriori (MAP) estimation problem:
\begin{equation}
\begin{aligned}
&\underset{\mathbf{x}_{k}\in\mathcal{X}}{\mathsf{maximize}}\,\,\, &\mathbb{P}(\mathbf{x}_{k}\vert\mathbf{y}_{1:k})\\
&\mathsf{subject\,\,to}  &\mathbf{g}(\mathbf{x}_{k})\leq\mathbf{0}
\end{aligned}
\label{eq:return-const mark-2-1}
\end{equation}
where $\mathbf{g}$ is a continuously differentiable function with a Lipschitz continuous gradient.

By applying Bayes' rule, we find that the posterior probability of state $\mathbf{x}_k$ given all available measurements $\mathbf{y}_{1:k}$ is proportional to
    \begin{equation}
        \mathbb{P}(\mathbf{x}_k\vert \mathbf{y}_{1:k})  \propto \mathbb{P}(\mathbf{x}_k\vert \mathbf{y}_{1:k-1})\mathbb{P}(\mathbf{y}_k\vert \mathbf{x}_{k}).
    \end{equation}
Therefore, the MAP estimate of  $\mathbf{x}_k$ can be obtained by solving the following problem:
\begin{equation}
\label{eq:log}
\begin{aligned}\underset{\mathbf{x}_{k}}{\mathsf{minimize}}\,\,\, & \quad-\log\mathbb{P}(\mathbf{x}_{k}\vert\mathbf{y}_{1:k-1})-\log\mathbb{P}(\mathbf{y}_{k}\vert\mathbf{x}_{k}).\\
\mathsf{subject\,\,to} & \quad \mathbf{g}(\mathbf{x}_{k})\leq\mathbf{0},\quad\mathbf{x}_{k}\in\mathcal{X}
\end{aligned}
\end{equation}

In the rest of the paper, the notation $\mathbf{x}_k$ is simplified to $\mathbf{x}$ for ease of representation. After simplification, the constrained state estimation problem with Student-t noise can be expressed as follows:
\begin{equation}
\begin{aligned}\underset{\mathbf{x}}{\mathsf{minimize}}\,\,\, & \quad F(\mathbf{x})\\
\mathsf{subject\,\,to} & \quad \mathbf{g}(\mathbf{x})\leq\mathbf{0}, \quad \mathbf{x}\in\mathcal{X}.
\end{aligned}
\tag{$\mathcal{P}$}
\label{prob_optim}
\end{equation}
Assume that, up to time $k-1$, the $\mathbb{P}(\mathbf{x}_k\vert \mathbf{y}_{1:k-1})$ is a Gaussian density $\mathcal{N}(\hat{\mathbf{x}}_{k\vert k-1}, \mathbf{P}_{k\vert k-1})$. The subsequent section will provide the reason for this assumption. Here, the objective function $F(\mathbf{x})$ is a nonconvex function defined as:
\begin{equation}
\begin{aligned}F(\mathbf{x})= & \lVert\mathbf{x}-\hat{\mathbf{x}}_{k\vert k-1}\rVert_{\mathbf{P}_{k\vert k-1}^{-1}}\\
 & +\sum_{i=1}^{n_{y}}(1+\nu_{i})\log\left(1+\frac{(\mathbf{C}_{i}\mathbf{x}-y_{k,i})^{2}}{\sigma_{i}^{2}\nu_{i}}\right),
\end{aligned}
\label{eq: F formula}
\end{equation}
where $\mathbf{C}_i$ denotes the row vector $\mathbf{C}[i,:]$. Problem \ref{prob_optim} is subject to more general constraints $g(\mathbf{x})\leq\mathbf{0}$, which can have different structures depending on the application and can be nonlinear and nonconvex. Examples of such constraints will be presented in Section \ref{subsec: constraints}.

The nonconvex nature of the objective function $F(\cdot)$ and the absence of assumed convexity in $\mathbf{g}(\cdot)$ pose challenges for solving Problem \ref{prob_optim}. Traditional optimization methods, such as the one proposed in \cite{qin2018new}, are designed to handle unconstrained problems and are not suitable for finding a solution to Problem \ref{prob_optim}. Additionally, other non-optimization based Student-t filters have not specifically addressed the reformulated Problem \ref{prob_optim}, which has resulted in the use of computationally expensive particle-based methods or less accurate estimation techniques.

\subsection{Examples of constraints $\mathbf{g}(\mathbf{x})\leq\mathbf{0}$ \label{subsec: constraints}}
In practical applications, various types of constraints can be imposed on the estimated state $\mathbf{x}_k$. Here, we provide examples of both linear and nonlinear constraints.

\subsubsection{Linear constraints } In portfolio modeling with state evolution, two linear constraints are defined as follows \cite{pizzinga2006state,markowitz1952portfolio}
    \begin{equation}
        \mathbf{1}^\top\mathbf{x}_k = 1,\quad \mathbf{x}_k \geq 0,
    \end{equation}
where $\mathbf{1}$ is a vector of ones. These constraints ensure that the portfolio weights sum up to one and that individual weights are non-negative.

\subsubsection{Nonlinear constraints}  In the case of constraining a vehicle to travel along a circular path with radius $r$, where the position is denoted by the two-dimensional state vector $\mathbf{x}_k = [{p}_x, {p}_y]^\top$, the circular route soft constraint can be formulated as \cite{straka2012truncation}:
    \begin{equation}
        (r-\epsilon)^2 \leq \mathbf{x}_k^\top\mathbf{x}_k \leq (r+\epsilon)^2,
        \label{Eqn: nonlinear constraint circle}
    \end{equation}
where $\epsilon > 0$. This constraint ensures that the estimated vehicle stays within a certain distance of the circular path.

In spacecraft tracking, the spacecraft orientation is represented by a quaternion $\mathbf{x}_k$, which must satisfy the quadratic constraint \cite{hutao2011rhc}:
    \begin{equation}
        \mathbf{x}_k^\top\mathbf{D}\mathbf{x}_k \leq 0,
    \label{Eqn: nonlinear constraint attitude}
    \end{equation}
 where $\mathbf{D}$ is an indefinite matrix \cite{hutao2011rhc}. This constraint ensures the validity of spacecraft orientation within a specified range.
 
  It is worth noting that some of the constraints above are nonconvex, further complicating the estimation of hidden states. In the subsequent sections, we present an approach that efficiently addresses the challenges.
\section{Proposed algorithm for state estimation}
To address the aforementioned requirements, we propose a novel constrained MM approach for solving the optimization problem described in \ref{prob_optim}. In this section, we begin by introducing the MM method and its application to \ref{prob_optim}. Subsequently, we present the complete algorithm in Kalman filter's framework.

\subsection{Proposed MM approach for Problem \ref{prob_optim}\label{subsec: proposed MM approach}}
The core idea of MM \cite{sun2016majorization} is to iteratively solve a series of subproblems instead of directly addressing the nonconvex problem using a nonconvex optimizer. Given the current point $\mathbf{x}^t$ at $t$-th iteration, we establish surrogate functions $\tilde{F}$, $\tilde{\mathbf{g}}$ for the objective function $F$ and the constraint function $\mathbf{g}$. Consequently, we formulate a subproblem as follows:

\begin{equation}
\begin{aligned}
    \underset{\mathbf{x}}{\mathsf{minimize}}\,\,\, &\quad  \tilde{F}\left(\mathbf{x};\mathbf{x}^{t}\right)\\
\mathsf{subject\,\,to} &\quad \Tilde{\mathbf{g}}\left(\mathbf{x};\mathbf{x}^t\right) \leq \mathbf{0} ,\quad \mathbf{x} \in \mathcal{X}.
\end{aligned}
\tag{$\mathcal{P}_s$}
\label{eq: mm subproblem}
\end{equation}
We require the $\tilde{F}$ and $\tilde{\mathbf{g}}$ are to be quadratic and strongly convex, making \ref{eq: mm subproblem} a Quadratic Constrained Quadratic Problem (QCCP). The choice of a QCQP subproblem is motivated by its convexity and the availability of efficient QCQP solvers to obtain optimal solutions. Through iterative resolution of \ref{eq: mm subproblem}, we aim to solve the constrained inference problem in \ref{prob_optim}.

\subsection{Selection of quadratic surrogate $\tilde{F}$ and $\tilde{\mathbf{g}}$\label{subsec: proposed f and g}}
In the following, we provide a detailed explanation of the construction of \ref{eq: mm subproblem}. According to \cite{sun2016majorization} and our requirements on surrogate function, $\tilde{F}$ should satisfy the following conditions \cite{sun2016majorization}:
\begin{enumerate}
    \item $\Tilde{F}(\mathbf{x}^t;\mathbf{x}^t) = F(\mathbf{y})$.
    \item $\Tilde{F}(\mathbf{x};\mathbf{x}^t) \geq F(\mathbf{x})$, $\forall \mathbf{x} \in \mathcal{X}$.
    \item $\Tilde{F}(\mathbf{x};\mathbf{x}^t)$ is continuous in $\mathbf{x}$ and $\mathbf{y}$.
    \item $\Tilde{F}(\mathbf{x};\mathbf{x}^t)$ is quadratic and strongly convex.
\end{enumerate}
Essentially, these conditions require $\tilde{F}$ to serve as a global upper bound that is convex and quadratic. Considering that the first term $\lVert\mathbf{x}-\hat{\mathbf{x}}_{k\vert k-1}\rVert_{\mathbf{P}_{k\vert k-1}^{-1}}$ 
 in Equation \eqref{eq: F formula} is already a convex quadratic term, our focus lies on constructing a quadratic upper bound for the nonconvex term in $F$, denoted as $F_{\text{ncvx}}$:
\begin{equation}
    F_{\text{ncvx}}(\mathbf{x}) \triangleq \sum_{i=1}^{n_{y}}(1+\nu_{i})\log\left(1+\frac{(\mathbf{C}_{i}\mathbf{x}-y_{k,i})^{2}}{\sigma_{i}^{2}\nu_{i}}\right).
\end{equation}
In the following parts, we propose three types of $\tilde{F}$ that fulfill the aforementioned conditions.

\subsubsection{Convexification of logarithm} The nonconvex nature of $F_{\text{ncvx}}$ stems from the logarithmic term. To address this, we can approximate the concave logarithm function using its first-order Taylor expansion around the current point,
$h_i(\mathbf{x}^{t})$:
    \begin{equation}
    \label{eq: log taylor}
    \log h_i\left(\mathbf{x}\right)\leq\log h_i\left(\mathbf{x}^{t}\right)+\frac{h_i\left(\mathbf{x}\right)-h_i\left(\mathbf{x}^{t}\right)}{h_i\left(\mathbf{x}^{t}\right)}, i = 1,\dots, n_y,
    \end{equation}
where 
\begin{equation}
    h_i(\mathbf{x}) = \left(1+\frac{(\mathbf{C}_{i}\mathbf{x}-y_{k,i})^{2}}{\sigma_{i}^{2}\nu_{i}}\right).
\end{equation}
Utilizing Equation \eqref{eq: log taylor}, we obtain the first surrogate function after some simplification:
\begin{equation}
\tilde{F}_\text{log}(\mathbf{x})=\lVert\mathbf{x}-\hat{\mathbf{x}}_{k\vert k-1}\rVert_{\mathbf{P}_{k\vert k-1}^{-1}}+\sum_{i=1}^{n_{y}}m_{i}^t\left(\mathbf{C}_{i}\mathbf{x}-y_{k,i}\right)^{2},
\label{eq: F_approx log}
\end{equation}
where $m_i^t$ is an introduced constant for brevity:
\begin{equation}
m_i^t = \frac{(1+\nu_i)}{\nu_i\sigma_i^2+\left(\mathbf{C}_i\mathbf{x}^{t} - y_{k,i}\right)^2}.
\end{equation}
\subsubsection{L-smooth based method}
The Descent Lemma \cite{bertsekas1997nonlinear} offers an alternative approach for constructing a surrogate function when the original function is Lipschitz smooth. We present the following lemma:

\begin{lemma}
     $F_{\text{ncvx}}(\mathbf{x})$ is Lipschitz smooth, leading to
    \begin{equation}
        F_{\text{ncvx}}(\mathbf{x}) \leq F_{\text{ncvx}}(\mathbf{y})+\nabla F_{\text{ncvx}} (\mathbf{y})^\top(\mathbf{x} - \mathbf{y})+ \frac{L}{2}\lVert \mathbf{x} -\mathbf{y}\rVert_2^2,
    \end{equation}
    with Lipschitz constant $L$ to be 
     \begin{equation}
            L = 2\sum_{i = 1}^{n_y} \frac{(\nu_i+1)}{\nu_i\sigma_i^2}\lVert \mathbf{C}_i\rVert_2^2.
        \end{equation}
    \label{lemma: Lipschitz gradient}
\end{lemma}
\begin{IEEEproof}
Suppose $\mathbf{z}_1,\ \mathbf{z}_2 \in \mathcal{X}^{n_x}$ are two variables. For brevity, we use $w_{1,i} = \mathbf{C}_i\mathbf{z}_1-y_{k,i}$, $w_{2,i}=\mathbf{C}_i\mathbf{z}_2-y_{k,i},\ i = 1,\dots,n_y$. By the triangle inequality, we have:
\begin{equation}
\begin{aligned}
    \lVert\nabla F_{\text{ncvx}}(\mathbf{z}_1) -&  \nabla F_{\text{ncvx}}(\mathbf{z}_2) \rVert_2 \\
     &\leq  \sum_{i = 1}^{n_x}\lVert \nabla f_{i}(\mathbf{z}_1) - \nabla f_{i}(\mathbf{z}_2)\rVert_2,
\end{aligned}
    \end{equation}
    \begin{equation}
            \begin{aligned}
        \lVert\nabla f_{2,i}(\mathbf{z}_1) - \nabla f_{2,i}(\mathbf{z}_2)\rVert_2  = 2(\nu_i+1)\lVert \mathbf{C}_i\rVert_2 l_{i}(\mathbf{z}_1,\mathbf{z}_2),
    \end{aligned}
    \end{equation}
        where 
        \begin{equation}
        \begin{aligned}
               l_{i}(\mathbf{z}_1,\mathbf{z}_2) & =\frac{\vert\nu_i\sigma_i^2\mathbf{C}_i(\mathbf{z}_1 - \mathbf{z}_2)+ \mathbf{C}_i(\mathbf{z}_1 - \mathbf{z}_2)w_1w_2\vert}{(\nu_i\sigma_i^2+w_1^2)(\nu_i\sigma_i^2+w_1^2)}\\&
                \leq \frac{1}{\nu_i\sigma_i^2}\lVert \mathbf{C}_i\rVert_2 \lVert \mathbf{z}_1-\mathbf{z}_2\rVert_2.
        \end{aligned}
    \end{equation}
\end{IEEEproof}
With this property, we introduce the second type of surrogate function as follows: 
    \begin{equation}
    \begin{aligned}
         &\Tilde{F}_\text{smooth}(\mathbf{x})=  \lVert\mathbf{x}-\hat{\mathbf{x}}_{k\vert k-1}\rVert_{\mathbf{P}_{k\vert k-1}^{-1}} \\ &+ \sum_{i = 1}^{n_x}2m_i^t(\mathbf{C}_i\mathbf{x}^{t} - y_{k,i})^\top\mathbf{C}_i\mathbf{x} 
         -L(\mathbf{x}^{t})^\top\mathbf{x}+ \frac{L}{2}\mathbf{x}^\top\mathbf{x}.
    \end{aligned}
    \label{eq: F_approx lipschitz}
    \end{equation}
    
So far, we have presented two types of surrogate functions for the objective function $F$. However, the constraints may exhibit diverse formats. Consequently, we propose a general approach to construct an approximation $\tilde{\mathbf{g}}$ when dealing with constraint $\mathbf{g}(\mathbf{x})\leq \mathbf{0}$ that are not convex quadratic. 
The surrogate function $\tilde{\mathbf{g}}$ should satisfy the following conditions \cite{marks1978general}:
\begin{enumerate}
    \item $\Tilde{\mathbf{g}}(\mathbf{x}^t;\mathbf{x}^t) = \mathbf{g}(\mathbf{x}^t)$.
    \item $\Tilde{\mathbf{g}}(\mathbf{x};\mathbf{x}^t) \geq \mathbf{g}(\mathbf{x})$, $\forall \mathbf{x} \in \mathcal{X}$.
    \item $\nabla\Tilde{\mathbf{g}}(\mathbf{x}^t;\mathbf{x}^t) =\nabla \mathbf{g}(\mathbf{x}^t) $.
    \item $\Tilde{\mathbf{g}}(\mathbf{x};\mathbf{x}^t)$ is differentiable and convex.
\end{enumerate}
To satisfy the above requirements, a quadratic surrogate function can be constructed as: 
\begin{equation}
\label{eq: surrogate for h}
\begin{split}
    \Tilde{\mathbf{g}}\left(\mathbf{x}; \mathbf{x}^{t}\right)&=  
    \mathbf{g}\left(\mathbf{x}^{t}\right) + \nabla \mathbf{g}\left( \mathbf{x}^{t}\right)^\top\left(\mathbf{x} - \mathbf{x}^{t}\right) +\frac{G}{2}\lVert \mathbf{x} -\mathbf{x}^t\rVert_2^2,
\end{split}
\end{equation}
where $G$ is the Lipschitz is the constant. 

Notice that by utilizing the surrogate functions presented in \eqref{eq: F_approx log} and \eqref{eq: F_approx lipschitz}, along with the quadratic approximation of the constraint in \eqref{eq: surrogate for h}, the subproblem \ref{eq: mm subproblem} transforms into a strongly convex QCQP problem.

\subsection{Implementation in the Kalman Filter Framework\label{subsec: proposed algorithm framework}}
In the previous subsections, our proposed method enables the estimation of the hidden state value at the current time step. However, to predict and estimate future state values, we would approximate $\mathbb{P}(\mathbf{x}_k\vert\mathbf{y}_{k\vert k-1})$ as a Gaussian distribution. Therefore, in this section, we would explain the reason for such approximation and also obtain the posterior estimation error covariance of $\mathbf{x}_k$, denoted as $\mathbf{P}_{k\vert k}$.

The traditional Kalman filter for a linear state-space model with independent univariate Gaussian observation noise $v_i \sim \mathcal{N}(0, r_i)$ can be equivalently formulated as solving the following optimization problem:
\begin{equation}
  \underset{\mathbf{x}_k\in\mathcal{X}}{\mathsf{minimize}}\quad \lVert \mathbf{x}_k - \hat{\mathbf{x}}_{k\vert k-1}\rVert_{\mathbf{P}_{k\vert k-1}^{-1}} 
    + \sum_{i=1}^{n_y} \frac{(\mathbf{C}_i\mathbf{x}_k - y_{k,i})^2}{r_{i}}.
\label{eq: Kalman MAP optimization problem}
\end{equation}
When comparing \ref{prob_optim} and Problem \eqref{eq: Kalman MAP optimization problem}, if we assume the constraint is inactive, i.e., the constraint $\mathbf{g}(\mathbf{x})\le\mathbf{0}$ is satisfied with strict inequality, then as indicated in \cite{agamennoni2012approximate}, the equivalent unconstrained Student-t filter can be regarded as an adaptive Kalman filter with a varying measurement noise covariance, i.e. $\mathbf{v}_k \sim \mathcal{N}(\mathbf{0},\mathbf{R}_k)$. It can also be seen from the surrogate function Equation \eqref{eq: F_approx log}, which is quadratic and $1/m_i^t$ can be viewed as the Gaussian noise variance. This simplification of the Student-t filter as a series of Kalman filters introduces the Gaussian posterior distribution, which possesses desirable mathematical properties, including closed-form solutions for propagation and updating steps in the inference process.

By adopting this simplification, we assume $\mathbf{R}_k = \text{diag}(r_{k,1},\dots, r_{k,n_y})$, and we show that our proposed filter can also be simplified as a series of adaptive Kalman filters with varying $r_{k,i}$ values given by:
\begin{equation}
\label{eq:rk}
r_{k,i}=\frac{\left(\mathbf{C}_{i}\hat{\mathbf{x}}_{k\vert k}-y_{k,i}\right)^{2}}{\left(1+\nu_{i}\right)\log\left(1+\frac{(\mathbf{C}_{i}\hat{\mathbf{x}}_{k\vert k}-y_{k,i})^{2}}{\sigma_{i}^{2}\nu_{i}}\right)}
\end{equation}
To summarize, our proposed Algorithm \ref{al:filter} differs from the Kalman filter in that it deals with state-space models with Student-t measurement noise. It faces a nonconvex constrained state estimation problem \ref{prob_optim} and requires iteratively solving QCQP subproblems.
\begin{algorithm}
    \caption{Proposed algorithm for constrained state-space models with Student-t noise\label{al:filter}} 
    \begin{flushleft} 
    \textbf{Input:} $\boldsymbol{\theta}$, $\mathbf{y}_{1:T}$
    \end{flushleft}
    \begin{algorithmic}[1]
    \FOR {$k = 1$ to $T$}
    \STATE $t \leftarrow 0$;
    \STATE $\hat{\mathbf{x}}_{k\vert k-1} = \mathbf{A}\hat{\mathbf{x}}_{k-1\vert k-1}$;
    \STATE $\mathbf{P}_{k\vert k-1} = \mathbf{A}^\top\mathbf{P}_{k-1\vert k-1}\mathbf{A}+\mathbf{Q}$;
    \STATE Let $\Tilde{\mathbf{x}}_k^0 \leftarrow \hat{\mathbf{x}}_{k\vert k-1}$;
    \REPEAT
    \STATE Construct surrogate functions $\tilde{F}$ and $\tilde{\mathbf{g}}$ given $\Tilde{\mathbf{x}}_{k}^{t}$;
    \STATE Solve the QCQP subproblem \ref{eq: mm subproblem} to obtain its optimal solution $\Tilde{\mathbf{x}}_{k}^{t+1}$;
    \STATE Let $t \leftarrow t+1$;
    \UNTIL{converge}
    \STATE Let $\hat{\mathbf{x}}_{k\vert k} \leftarrow \Tilde{\mathbf{x}}_{k}^{t+1}$;
    \STATE Calculate $\mathbf{R}_k= \text{diag}(r_{k,1},\dots, r_{k,n_y})$ according to \eqref{eq:rk};
    \STATE $\mathbf{K}_k = \mathbf{P}_{k\vert k-1}\mathbf{C}^\top(\mathbf{C}\mathbf{P}_{k\vert k-1}\mathbf{C}^\top+\mathbf{R}_k)^\top$;
    \STATE $\mathbf{P}_{k\vert k} = \mathbf{P}_{k\vert k-1} - \mathbf{K}_k\mathbf{C}\mathbf{P}_{k\vert k-1}$ 
    \ENDFOR
    \end{algorithmic}
    \begin{flushleft}
    \textbf{Output:} $\hat{\mathbf{x}}_{k\vert k}$, $\mathbf{P}_{k\vert k}$, $k = 1,\dots,T$
    \end{flushleft}
\end{algorithm}

\section{Experiments}
To assess the performance of the proposed method, two experiments are conducted. The first experiment aims to evaluate the robustness of the proposed framework under unconstrained conditions. In the second experiment, we focus on assessing the performance of the proposed algorithm under constrained conditions. 
\subsection{Experiment 1}
In this experiment, a two-state system is considered, with the state transition matrix and measurement matrix given by:
\begin{equation}
\mathbf{A}=\left[\begin{array}{cc}
\cos\left(0.2\pi\right) & \sin\left(0.2\pi\right)\\
-\sin\left(0.2\pi\right) & \cos\left(0.2\pi\right)
\end{array}\right],\quad
\mathbf{C}=\left[\begin{array}{cc}
1 & 0\\
0 & 1
\end{array}\right].\nonumber
\end{equation}
The state noise covariance matrix was set as $\mathbf{Q} = \sigma^2\mathbf{I}$, where $\sigma^2=0.1$. And we assume a contaminated Gaussian noise pattern for the measurement noise:

\begin{equation}
\label{eq:exp1 noise}
v_{t}\sim\begin{cases}
\mathcal{N}(0,0.1) & \text{with probability}\ 0.9;\\
\mathcal{N}(0,10) & \text{with probability}\ 0.1.
\end{cases}
\end{equation}
We compare our methods with the Kalman Filter (KF), VB based Student-t filter (TFVB) \cite{huang2017novel}, Majorization Minimization based Laplace Kalman filter (LFMM) \cite{wang2017laplace} and particle filter (PF). PF assumes the Student-t measurement noise and $10,000$ particles, implemented using the R package $\mathsf{pomp}$.
The two proposed algorithms in this study are denoted as unconstrained and TFMM-smooth corresponding to approximations \eqref{eq: F_approx log} and \eqref{eq: F_approx lipschitz}. The degree of freedom for TFVB and the proposed filters is set to $3$. And the variances of all filters are set to the variance of $v_t$.
The root mean squared error (RMSE) is utilized as a metric, which is defined as follows:
\begin{equation}
    \text{RMSE} = \sqrt{\frac{1}{T}\sum_{k = 1}^T(\hat{\mathbf{x}}_{k\vert k} -\mathbf{x}_k^{\text{true}})^\top(\hat{\mathbf{x}}_{k\vert k} -\mathbf{x}_k^{\text{true}} )}.
\end{equation}
For each experiment, $1000$ data points are generated, and the experiment is repeated for $100$ runs. The implementations of all methods are carried out in R and executed on 2.10GHz Intel Core i7-12700 machines. 

\begin{figure}[]
    \centering
    \includegraphics[width=0.8\linewidth]{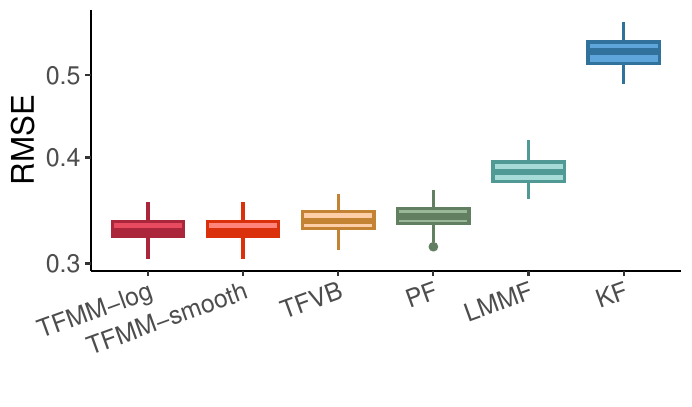}
    \caption{RMSE of proposed methods and benchmarks.}
    \label{fig:exp1 rmse}
\end{figure}

Figure \ref{fig:exp1 rmse} displays the RMSE values of the methods. It is evident that all the robust filters outperform the traditional KF. TFVB algorithm performs slightly worse than the proposed method, as it employs the mean-field approximation. PF does not exhibit better performance than the TFVB and proposed methods in this case, as its performance is sensitive to the choice of importance distribution and the accuracy of the propagation step. This observation aligns with similar findings in the \cite{nurminen2018skew}.

Table \ref{table: exp1} provides the average CPU times for the methods. It is evident that the proposed method achieves higher computational efficiency compared to TFVB and LFMM. Notably, the PF takes approximately 4000 times longer than the proposed methods. In summary, the proposed method not only outperforms the other robust filters in terms of accuracy but also demonstrates high computational efficiency.
\begin{table}[]
\centering
\caption{Average CPU time of proposed methods and benchmarks (unit: second)\label{table: exp1}}
\begin{tabular}{|c|c|c|c|c|c|}
\hline 
\textbf{TFMM-log} & \textbf{TFMM-smooth }  & \textbf{TFVB}&\textbf{LFMM} &  \textbf{KF} & \textbf{PF}\tabularnewline
\hline 
1.53 & 1.81 & 2.29&9.09 & 0.37 &  8,717\tabularnewline
\hline 
\end{tabular}
\end{table}
\subsection{Experiment 2}
In the second simulation example, we consider the movement of a ground vehicle along a circular road segment \cite{yang2009kalman}. The origin of the x-y coordinates is chosen as the turn center, with a turning radius of $r = 100\text{m}$. The vehicle maintains a constant linear speed of $4\text{m/s}$. The state is defined as $\mathbf{x}_k = [p_x, v_x, p_y,v_y]^\top$, where $p_x, p_y$ are the position and $v_x, v_y$ are the velocity. The initial state is $\mathbf{x}_{1}^{\text{true}} = \left[0 \text{m},4\text{m/s},100\text{m},0\text{m/s}\right]$. The vehicle's position is tracked using the following measurement equation:
\begin{equation}
    \mathbf{y}_k = \left[\begin{array}{cccc}
1 & 0 &0 &0\\
0&0&1&0
\end{array}\right]\mathbf{x}_k + \mathbf{v}_k,
\end{equation}

The sensor utilizes a discrete-time second-order kinematic model to describe the vehicle's motion:
\begin{equation}
    \mathbf{x}_k =  \left[\begin{array}{cccc}
1 & T &0 &0\\
0&1&0&0\\
0&0&1&T\\
0&0&0&1
\end{array}\right]\mathbf{x}_{k-1} +  \left[\begin{array}{cc}
0.5T^2 &0\\
T&0\\
0&0.5T^2\\
0&T
\end{array}\right]\mathbf{w}_k,
\end{equation}
where $T = 1$ and $\mathbf{w}_k \sim \mathcal{N}(0, \mathbf{Q})$. In the simulation, the covariance matrix $\mathbf{Q} = \mathbf{I}\sigma_q^2$ is used, where $\sigma_q^2 = 1.5$. The initial state is selected to be the same as the true state, and the initial estimation error covariance is selected to be $\mathbf{P}_1 = \mathbf{I}\sigma_p^2$, where $\sigma_p^2 = 1$.

Given the characteristics of the vehicle's track, it is reasonable to impose the constraints on $p_x, p_y$ of each time step $k$. We denote $p_x$ and $p_y$ at time $k$ as $\mathbf{x}_{k,1}$ and $ \mathbf{x}_{k,3}$. It is expected that $\mathbf{x}_{k,1}^2 + \mathbf{x}_{k,3}^2$ closely approximates $r^2$s. To account for this, we introduce the following constraint:
\begin{equation}
   (r-0.1)^2 \leq \mathbf{x}_{k,1}^2 + \mathbf{x}_{k,3}^2 \leq (r+0.1)^2.
\end{equation}
This experiment aims to demonstrate two key concepts: the significance of incorporating constraints into the estimation process and the superior capability of our proposed method. To achieve this, we test the following algorithms:
\begin{itemize}
    \item Our proposed algorithm incorporates constraints into the estimation process, denoted as constrained (prop.), 
    \item Our proposed unconstrained filter serves as a baseline for comparison, denoted as unconstrained. 
    \item The projection-based method \cite{yang2009kalman} serves as a benchmark for constrained method, denoted as constrained (bech.). This method projects the unconstrained estimation onto a specific surface defined by the constraints:
    \begin{equation}
    \begin{aligned}\underset{\mathbf{x}}{\mathsf{minimize}}\,\,\, &\quad  (\hat{\mathbf{x}}_{k\vert k}-\mathbf{x})^\top\mathbf{P}_{k\vert k}^{-1}(\hat{\mathbf{x}}_{k\vert k}-\mathbf{x}), \\
\mathsf{subject\,\,to} &\quad  (r-0.1)^2 \leq \mathbf{x}_{1}^2 + \mathbf{x}_{3}^2 \leq (r+0.1)^2.\\
\end{aligned}
\end{equation}
\end{itemize} 
The degrees of freedom for all filters are $\nu_1 = \nu_2 = 3$, and the variances of all filters are set to the variance of $v_t$.
\begin{figure}
\begin{centering}
\subfloat[RMSE of position estimation]{\includegraphics[width=0.5\linewidth]{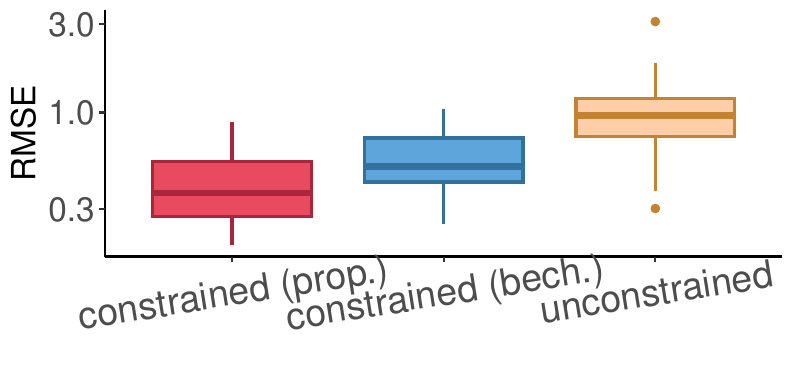}}
\subfloat[RMSE of velocity estimation]{\includegraphics[width=0.5\linewidth]{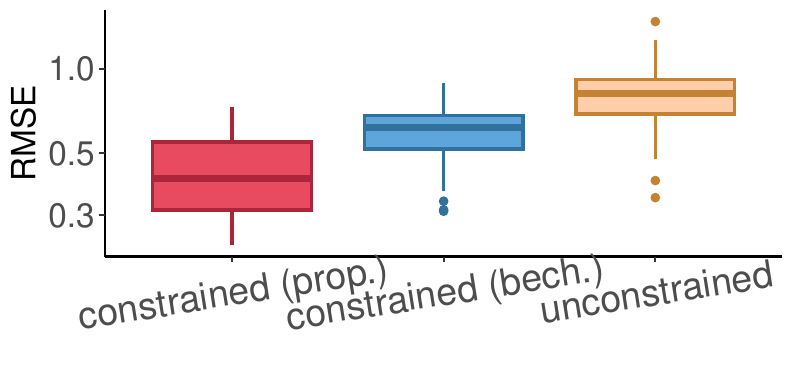}}
\par\end{centering}
\caption{RMSE of position and velocity estimation.}
\label{fig:exp2-posvol}
\end{figure}
\begin{figure}
    \centering
    \includegraphics[width=0.9\linewidth]{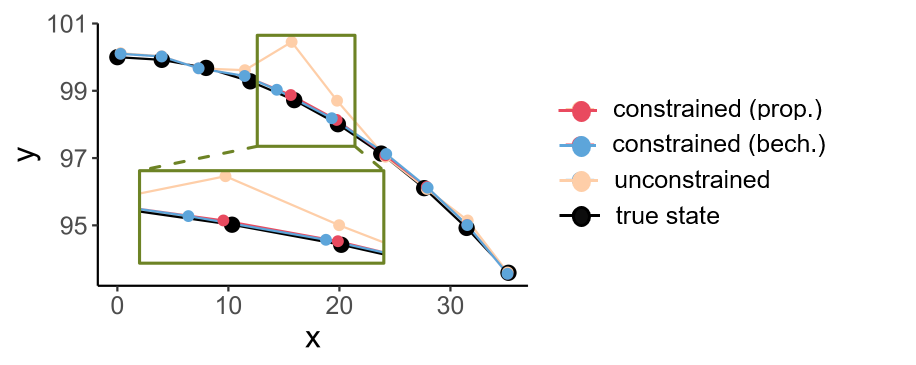}
    \caption{Path estimation for a specific road segment.}
    \label{fig:path}
\end{figure}
\begin{figure}
    \centering
    \includegraphics[width=0.9\linewidth]{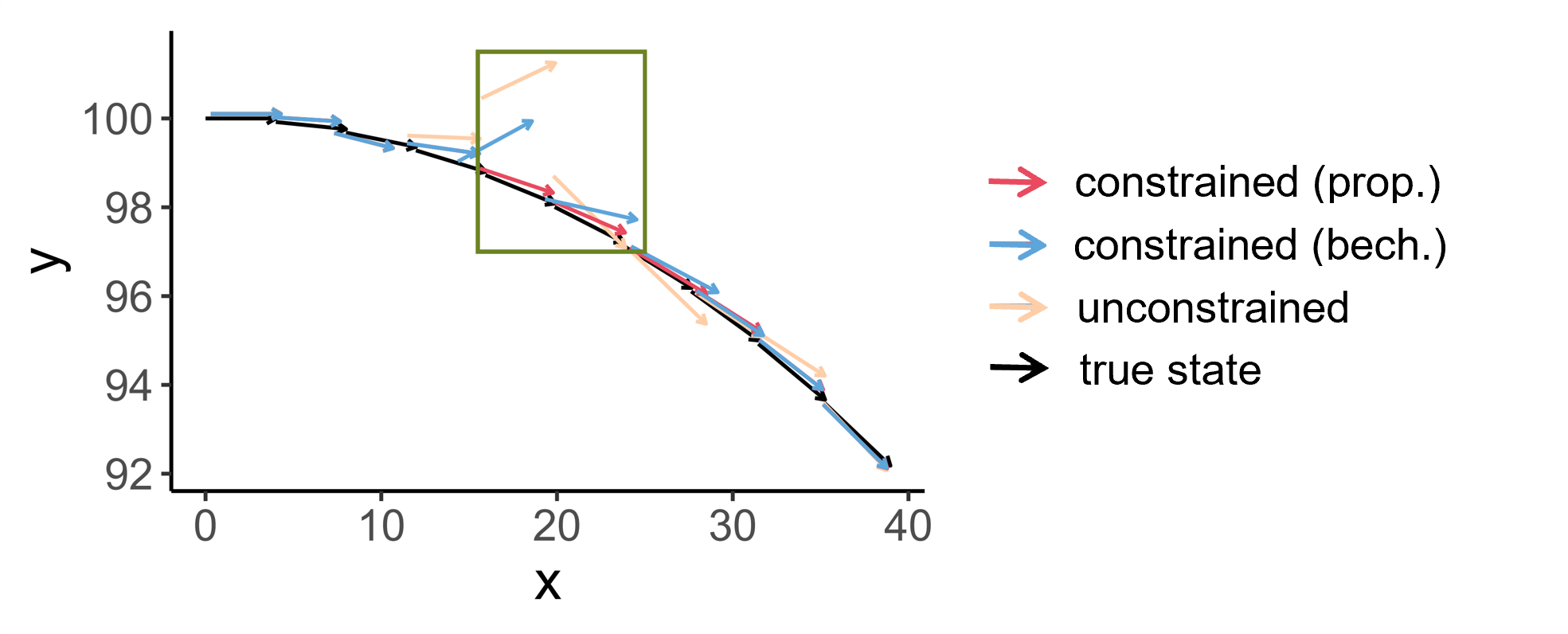}
    \caption{Velocity estimation for a specific road segment.}
    \label{fig:vel}
\end{figure}
We generate $35$ samples for each road segment and run the experiment for $50$ times. The RMSE is calculated for both position and velocity. The performance of the filters is shown in Figure \ref{fig:exp2-posvol}. The results show that the proposed method, constrained (prop.), exhibits superior performance compared to the other methods in position and velocity estimation. 

Figure \ref{fig:path} shows the part of the trajectory obtained from a particular experiment. It is evident that in some time instances, the estimation of unconstrained deviates from the true road. While constrained (bech.) successfully projects unconstrained estimation on to the true road, but the position estimation is less precise than the results obtained with constrained (prop.).  

Figure \ref{fig:vel} presents the velocity profile obtained from a specific experiment. The arrows in  Figure \ref{fig:vel} denote the velocity vectors. It is noticeable that constrained (bech.) does not consistently rectify the velocity estimation obtained by unconstrained. In contrast, our proposed method, constrained (prop.), estimates the velocity accurately.

\section{Conclusion}
We have proposed a robust state estimation algorithm using the Student-t distribution for modeling heavy-tailed noise. The approach employs the Majorization-minimization (MM) technique to transform the nonconvex objective function into a quadratic form. Our method also accommodates various constraints. Numerical results demonstrate the effectiveness of our approach in achieving robust and constrained estimation.

\bibliographystyle{IEEEtran}
\bibliography{paper_YuXiuPalomar}

\begin{thebibliography}{10}
\providecommand{\url}[1]{#1}
\csname url@samestyle\endcsname
\providecommand{\newblock}{\relax}
\providecommand{\bibinfo}[2]{#2}
\providecommand{\BIBentrySTDinterwordspacing}{\spaceskip=0pt\relax}
\providecommand{\BIBentryALTinterwordstretchfactor}{4}
\providecommand{\BIBentryALTinterwordspacing}{\spaceskip=\fontdimen2\font plus
\BIBentryALTinterwordstretchfactor\fontdimen3\font minus \fontdimen4\font\relax}
\providecommand{\BIBforeignlanguage}[2]{{%
\expandafter\ifx\csname l@#1\endcsname\relax
\typeout{** WARNING: IEEEtran.bst: No hyphenation pattern has been}%
\typeout{** loaded for the language `#1'. Using the pattern for}%
\typeout{** the default language instead.}%
\else
\language=\csname l@#1\endcsname
\fi
#2}}
\providecommand{\BIBdecl}{\relax}
\BIBdecl

\bibitem{franses2008simple}
P.~H. Franses, M.~Van Der~Leij, and R.~Paap, ``A simple test for {GARCH} against a stochastic volatility model,'' \emph{Journal of Financial Econometrics}, vol.~6, no.~3, pp. 291--306, 2008.

\bibitem{zhu2013variational}
H.~Zhu, H.~Leung, and Z.~He, ``A variational {Bayesian} approach to robust sensor fusion based on {Student-t} distribution,'' \emph{Information Sciences}, vol. 221, pp. 201--214, 2013.

\bibitem{teixeira2009state}
B.~O. Teixeira, J.~Chandrasekar, L.~A. T{\^o}rres, L.~A. Aguirre, and D.~S. Bernstein, ``State estimation for linear and non-linear equality-constrained systems,'' \emph{International Journal of Control}, vol.~82, no.~5, pp. 918--936, 2009.

\bibitem{kyriakides2005multiple}
I.~Kyriakides, D.~Morrell, and A.~Papandreou-Suppappola, ``Multiple target tracking with constrained motion using particle filtering methods,'' in \emph{1st IEEE International Workshop on Computational Advances in Multi-Sensor Adaptive Processing}, 2005, pp. 85--88.

\bibitem{arulampalam2002tutorial}
M.~S. Arulampalam, S.~Maskell, N.~Gordon, and T.~Clapp, ``A tutorial on particle filters for online nonlinear/non-{Gaussian} {Bayesian} tracking,'' \emph{IEEE Transactions on signal processing}, vol.~50, no.~2, pp. 174--188, 2002.

\bibitem{li2006t}
S.~Li, H.~Wang, and T.~Chai, ``A t-distribution based particle filter for target tracking,'' in \emph{2006 American Control Conference}, 2006, pp. 2191--2196.

\bibitem{xu2013robust}
D.~Xu, C.~Shen, and F.~Shen, ``A robust particle filtering algorithm with non-{Gaussian} measurement noise using {Student}-t distribution,'' \emph{IEEE Signal Processing Letters}, vol.~21, no.~1, pp. 30--34, 2013.

\bibitem{bishop2006pattern}
C.~M. Bishop and N.~M. Nasrabadi, \emph{Pattern recognition and machine learning}.\hskip 1em plus 0.5em minus 0.4em\relax New York, USA: Springer, 2006.

\bibitem{wang2017laplace}
H.~Wang, H.~Li, W.~Zhang, and H.~Wang, ``Laplace {${\ell}$1} robust {Kalman} filter based on majorization minimization,'' in \emph{2017 20th International Conference on Information Fusion (Fusion)}, 2017, pp. 1--5.

\bibitem{wen1992model}
W.~Wen and H.~F. Durrant-Whyte, ``Model-based multi-sensor data fusion,'' in \emph{Proceedings 1992 IEEE international conference on robotics and automation}, 1992, pp. 1720--1726.

\bibitem{porrill1988optimal}
J.~Porrill, ``Optimal combination and constraints for geometrical sensor data,'' \emph{The International journal of robotics research}, vol.~7, no.~6, pp. 66--77, 1988.

\bibitem{simon2002kalman}
D.~Simon and T.~L. Chia, ``Kalman filtering with state equality constraints,'' \emph{IEEE Transactions on Aerospace and Electronic Systems}, vol.~38, no.~1, pp. 128--136, 2002.

\bibitem{amor2018constrained}
N.~Amor, G.~Rasool, and N.~C. Bouaynaya, ``Constrained state estimation--a review,'' \emph{arXiv preprint arXiv:1807.03463}, 2018.

\bibitem{straka2012truncation}
O.~Straka, J.~Dun{\'\i}k, and M.~{\v{S}}imandl, ``Truncation nonlinear filters for state estimation with nonlinear inequality constraints,'' \emph{Automatica}, vol.~48, no.~2, pp. 273--286, 2012.

\bibitem{benidis2018optimization}
K.~Benidis, Y.~Feng, D.~P. Palomar \emph{et~al.}, ``Optimization methods for financial index tracking: From theory to practice,'' \emph{Foundations and Trends{\textregistered} in Optimization}, vol.~3, no.~3, pp. 171--279, 2018.

\bibitem{qin2018new}
F.~Qin, L.~Chang, and F.~Zha, ``New look at the {Student's} t-based {Kalman} filter from maximum a posterior perspective,'' \emph{IET Radar, Sonar \& Navigation}, vol.~12, no.~8, pp. 795--800, 2018.

\bibitem{pizzinga2006state}
A.~Pizzinga and C.~Fernandes, ``State space models for dynamic style analysis of portfolios,'' \emph{Brazilian Review of Econometrics}, vol.~26, no.~1, pp. 31--66, 2006.

\bibitem{markowitz1952portfolio}
H.~M. Markowitz, ``Portfolio selection,'' \emph{Journal of Finance}, vol.~7, no.~1, pp. 77--91, 1952.

\bibitem{hutao2011rhc}
C.~Hutao, C.~Xiaojun, X.~Rui, and C.~Pingyuan, ``{RHC-based} attitude control of spacecraft under geometric constraints,'' \emph{Aircraft Engineering and Aerospace Technology}, vol.~83, no.~5, pp. 296--305, 2011.

\bibitem{sun2016majorization}
Y.~Sun, P.~Babu, and D.~P. Palomar, ``Majorization-minimization algorithms in signal processing, communications, and machine learning,'' \emph{IEEE Transactions on Signal Processing}, vol.~65, no.~3, pp. 794--816, 2016.

\bibitem{bertsekas1997nonlinear}
D.~P. Bertsekas, ``Nonlinear programming,'' \emph{Journal of the Operational Research Society}, vol.~48, no.~3, pp. 334--334, 1997.

\bibitem{marks1978general}
B.~R. Marks and G.~P. Wright, ``A general inner approximation algorithm for nonconvex mathematical programs,'' \emph{Operations research}, vol.~26, no.~4, pp. 681--683, 1978.

\bibitem{agamennoni2012approximate}
G.~Agamennoni, J.~I. Nieto, and E.~M. Nebot, ``Approximate inference in state-space models with heavy-tailed noise,'' \emph{IEEE Transactions on Signal Processing}, vol.~60, no.~10, pp. 5024--5037, 2012.

\bibitem{huang2017novel}
Y.~Huang, Y.~Zhang, N.~Li, Z.~Wu, and J.~A. Chambers, ``A novel robust {Student's} t-based {Kalman} filter,'' \emph{IEEE Transactions on Aerospace and Electronic Systems}, vol.~53, no.~3, pp. 1545--1554, 2017.

\bibitem{nurminen2018skew}
H.~Nurminen, T.~Ardeshiri, R.~Pich{\'e}, and F.~Gustafsson, ``Skew-$ t $ filter and smoother with improved covariance matrix approximation,'' \emph{IEEE Transactions on Signal Processing}, vol.~66, no.~21, pp. 5618--5633, 2018.

\bibitem{yang2009kalman}
C.~Yang and E.~Blasch, ``Kalman filtering with nonlinear state constraints,'' \emph{IEEE Transactions on Aerospace and Electronic Systems}, vol.~45, no.~1, pp. 70--84, 2009.

\end{thebibliography}

\end{document}